\documentclass{article}

\usepackage{arxiv}

\usepackage[utf8]{inputenc} 
\usepackage[T1]{fontenc}    
\usepackage{hyperref}       
\usepackage{url}            
\usepackage{booktabs}       
\usepackage{amsfonts}       
\usepackage{nicefrac}       
\usepackage{microtype}      
\usepackage{lipsum}
\usepackage{graphicx}

\title{Backronym}

\author{
  Arip Asadulaev \\
  INFORNOPOLITAN \\
  \texttt{aripdotcom@mail.ru} \\
}

\begin{document}
\maketitle

\begin{abstract}
The field of Machine Learning research is divided into subject areas, where each area tries to solve a specific problem, using specific methods. In recent years, borders have almost been erased, and many areas inherit methods from other areas. This trend leads to better results and the number of papers in the field is growing every year. The problem is that the amount of information is also growing, and many methods remain unknown in a large number of papers. In this work, we propose the concept of inheritance between machine learning models, which allows conducting research, processing much less information, and pay attention to previously unnoticed models. We hope that this project will allow researchers to find ways to improve their ideas. In addition, it can be used by researchers to publish their methods too. Project is available by link: https://www.infornopolitan.xyz/backronym
\end{abstract}

\keywords{Machine Learning \and 3D visualization}

\section{Motivation}
Machine Learning researchers analyzing hundreds of papers every year, and it is rather difficult to structure such type of information. On average, the most valuable information extracted from a Machine Learning paper is method components and their interaction. To remember all processed methods is problematic, and it is necessary to somehow save it.

It is important to note that today, almost any new method is the development of an older method or compilation of a set of methods.  To represent such type of information as a table or a list it not optimal. In our opinion, a strictly fixed graph, with edges between methods and their components is the best way to do it. 

We thought about cite graph and unfortunately, the citation does not reflect very well which methods are directly used in the architecture. We decided to hand annotate it and write down from which components every method consists, to give an opportunity to researchers see preprocessed info about methods.

During papers analyzing, we realized how many methods with excellent results are still not used, due to the fact that they were lost among many other papers. There are a lot of models being developed, and often research's on a new conference do not inherit most of the improvements proposed a year ago.

There are many amazing services that allow you to monitor the emergence of new articles in the field. But all these methods were aimed primarily at the acquisition of knowledge, but not for help in research. Our platform aims to diversify research. Creativity is more important than the experience, so the tools that can help you to be more creative should be an integral part of the research process.

We have tools that make it easy to conduct experiments, run and evaluate models, but we do not have methods that allow us to extend our model in a ideas level. In Machine learning, and Artificial Intelligence research as a whole, fresh and elegant ideas play a key role.

Other fields of science cannot boast of such tools too, but we are absolutely sure that Machine Learning and other areas that solve difficult problems, needs an extraordinary approach for ideas improvements.

\section{The Graph}
In the graph each paper can be presented as one single method, for example, Autoencoder (AE), or separately, Autoencoder (AE) -> Encoder (ENCDR), Decoder (DCDR). Models inherit from each other but, some method may not use another method fully, but only some part of it. For example, GAN[1] consist of Generator(GEN), Discriminator (DIS), Adversarial Autoencoder (AAE)[2] based on Autoencoder (AE)[3] and DIS. Models have the edge to other methods, if it based on this method or include it directly in architecture, for example for AAE we will have an edge to AE and DIS. We call this graph BACKRONYM. 

In just a few months, we took out about ~250 articles from the NeurIPS 2019 conference and ~250 other papers on which they are based.
We preprocess papers and presented info in a matrix with 10 columns:
\begin{itemize}
    \item Paper title
    \item Link to paper
    \item Names of authors
    \item Release date
    \item Place of publication
    \item Method name
    \item Subject area (Using For)
    \item Acronym of method name
    \item A brief description of the method
    \item "Based on":the list of methods on which this method is based (list of acronyms which are available in table).
\end{itemize}
The graph is built using "Based on" column, where each row consists of a list of methods on which the method is based, Fig. 1.

The way we analyzed papers is very far from ideal. Most of the areas were completely unfamiliar to us, to understand them took several days. Sometimes we could not find the right description for some methods, and the abstract of the article was used for this. The graph is interactive, clicking on the node opens a list with meta-information from columns, Fig. 2.

The “Subject area (Using For)” column allows creating subgraphs with inheritances inside the field, where the method is marked in red if it using in other areas, see Fig. 3.

The visualization is built using the 3d-force-graph javascript library: https://github.com/vasturiano/3d-force-graph 

Now the simplest way to use the graph it skips connections. For example, you use CNN[4] in your model, and in a graph, you can see another method that inherited from CNN. You can make little research and try just to replace your CNN with the advanced version.

It is very difficult to create an inheritance graph that would be fully consistent with the truth. Based on this, we created an opportunity for authors to add their own methods and make changes to existing ones in the graph. After all, no one except the authors of the article knows how to disassemble and describe their method in the best way. https://www.infornopolitan.xyz/add-research 

\section{Discussion and Future Work}

Nevertheless, We believe that the main thing that the graph can give is associations. We were excited to know how GAN idea occurred to Ian Goodfellow. Talking on Artificial Intelligence podcast by Lex Fridman[5], Ian said that it was motivated by Boltzmann machine “positive” and “negative” training phases. This is a great example when a model inherits the properties of another not directly but very abstractly.

We would like very much to see this type of connections in this graph, and it depends solely on authors of papers, will or not they share what was motivated them to create such type of model to solve some problem. We think that the story not only about what methods were used specifically in their architecture but also ideologically on which concepts and knowledge the solution was formed, can allow us to be much more resourceful.

This project is community-driven. We want to make it better and motivate more peoples to add their models to the graph. More accurate information about methods and better visualization technologies can really make it a very useful tool.

In the future, with community support, we can level up BACKRONYM and built the system which will recommend us how to extend research. For this scenario, we plan to give to users the ability to visualize their own knowledge, publicly or privately, irrespective of the main graph. For example, user can build a graph of all methods which he knows or even graph of one model components, and the system will recommend him the most useful paper or method. Also, we plan to add: 1) 2D visualization. 2) Ability to build and save individual subgraphs. 3) Search by method name.

It seems that even today this graph take a place to be because it probably allows someone to get associations that may help to create some new method or extend current. Today, it’s just a graph, but tomorrow we will have the tools to objectively get machine recommendations and automatically evaluate the impact the proposed idea on the way to General Machine Intelligence systems.
\bibliographystyle{unsrt}  


\section{Images}

\begin{figure}[h!]
  \centering
  \includegraphics[scale=0.40]{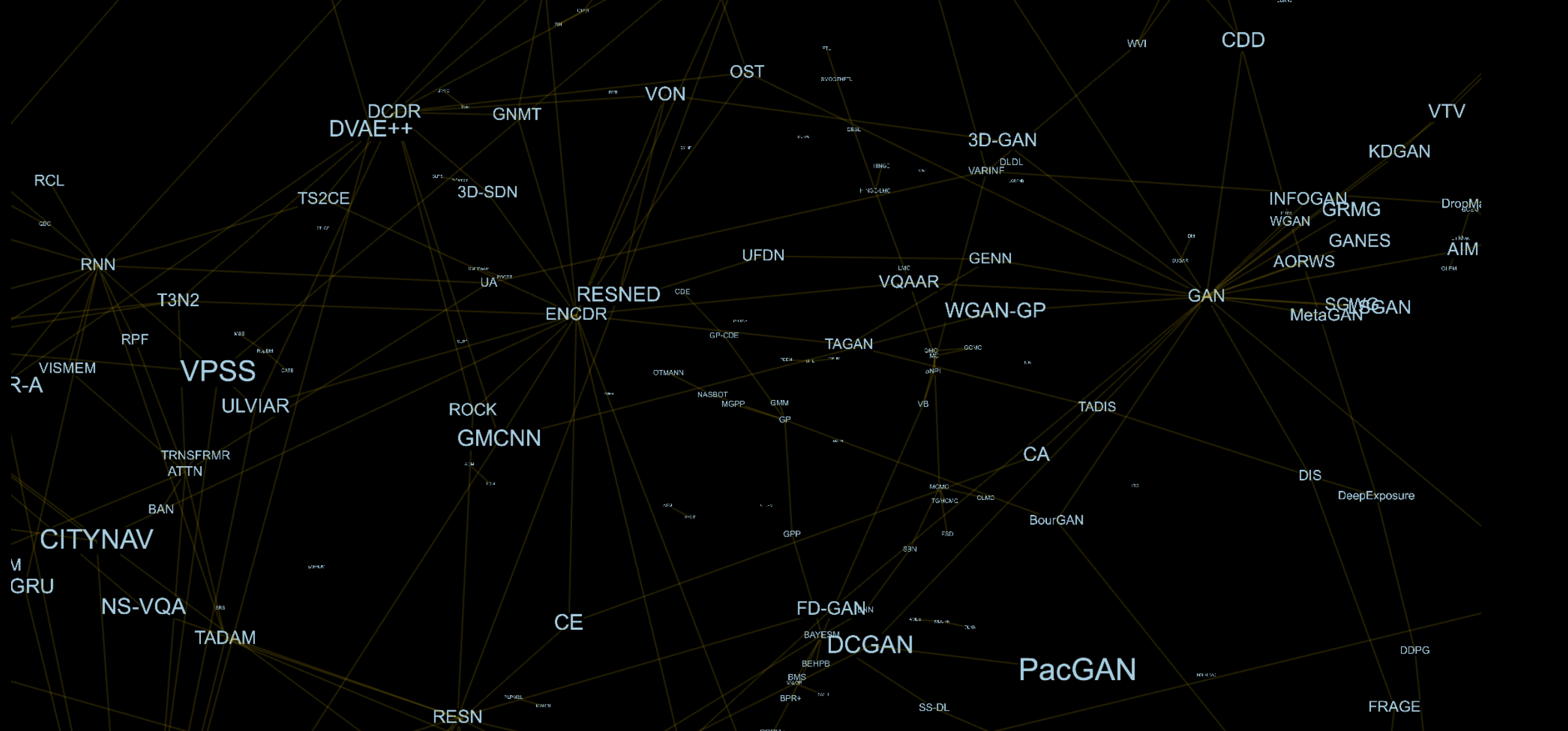}
  \caption{}
  \label{fig:fig1}
\end{figure}

\begin{figure}
  \centering
  \includegraphics[scale=0.43]{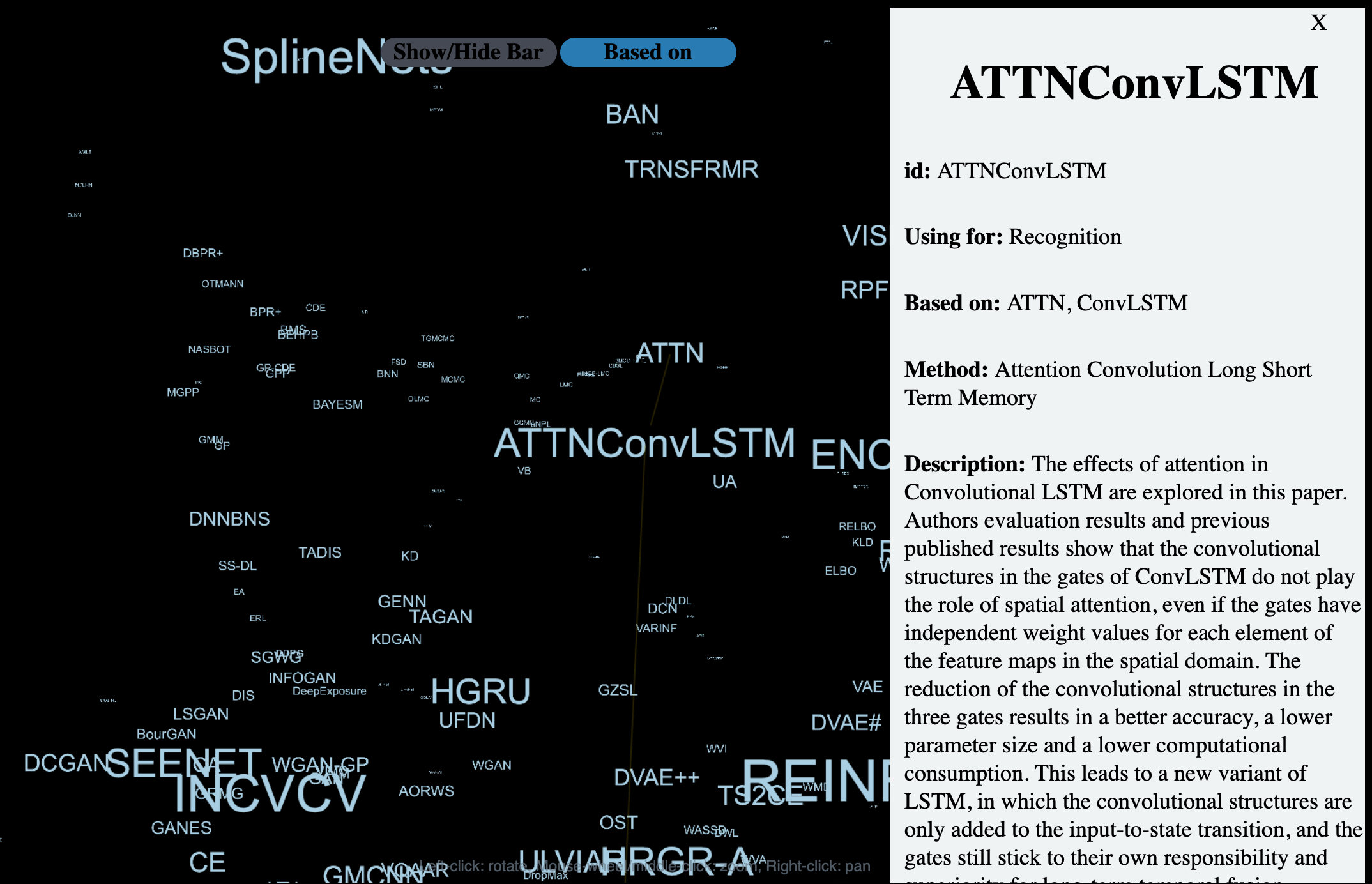}
  \caption{}
  \label{fig:fig2}
\end{figure}

\begin{figure}
  \centering
  \includegraphics[scale=0.43]{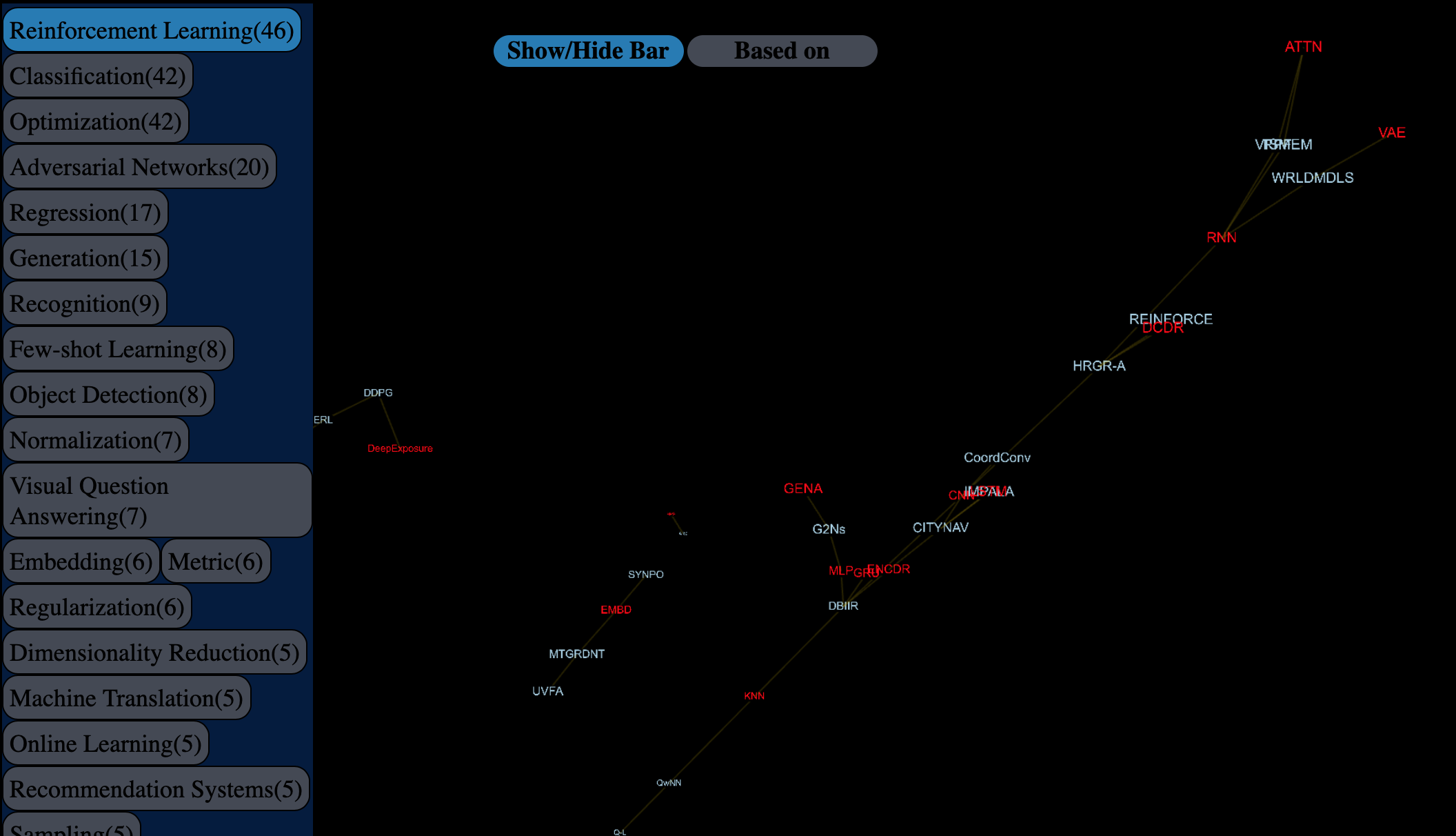}
  \caption{}
  \label{fig:fig3}
\end{figure}

\end{document}